 \newlength\smallfigwidth
\documentclass[aps,prb,twocolumn,floatfix,showpacs,amsmath,
amssymb ] {
revtex4-1}

\usepackage{amsmath}    
\usepackage{graphicx}   
\usepackage{verbatim}   
\usepackage{color}      
\usepackage{subfigure}  
\usepackage{float}
\usepackage{hyperref}   

\begin{document}

\preprint{UFES}

\title{Realization of magnetic monopoles current in an artificial spin ice device: A step towards magnetronics}
\author{R.\ P.\  Loreto}
\email{renan.loreto@gmail.com} \affiliation{ Departamento de F\'{i}sica,
Universidade Federal de Vi\c cosa, Vi\c cosa, 36570-900, Minas Gerais, Brazil }
\author{L.\ A.\  Morais}
\email{leonardoassisfisica@gmail.com} \affiliation{ Departamento de F\'{i}sica,
Universidade Federal de Vi\c cosa, Vi\c cosa, 36570-900, Minas Gerais, Brazil }
\author{R.\ C.\ Silva}
\email{rcs.fisica@gmail.com} \affiliation{ Departamento de F\'{i}sica,
Universidade Federal de Vi\c cosa, Vi\c cosa, 36570-900, Minas Gerais, Brazil }
\author{F.\ S.\ Nascimento}
\email{fabiosantos.ba@gmail.com} \affiliation{ Departamento de F\'{i}sica,
Universidade Federal de Vi\c cosa, Vi\c cosa, 36570-900, Minas Gerais, Brazil }
\author{C.\ I.\ L.\ Araujo}
\email{dearaujo@ufv.br} \affiliation{ Departamento de F\'{i}sica,
Universidade Federal de Vi\c cosa, Vi\c cosa, 36570-900, Minas Gerais, Brazil }
\author{L.\ A.\ S.\ M\'{o}l}
\email{lucasmol@yahoo.com} \affiliation{ Departamento de F\'{i}sica,ICEx, Universidade Federal de Minas Gerais, Belo Horizonte 31270-901, Minas Gerais, Brazil}
\author{W.\ A.\  Moura-Melo}
\email{winder@ufv.br} \affiliation{ Departamento de F\'{i}sica,
Universidade Federal de Vi\c cosa, Vi\c cosa, 36570-900, Minas Gerais, Brazil }
\author{A.\ R.\  Pereira}
\email{apereira@ufv.br} \affiliation{ Departamento de F\'{i}sica,
Universidade Federal de Vi\c cosa, Vi\c cosa, 36570-900, Minas Gerais, Brazil }
\date{April 10, 2014}

\begin{abstract}
Magnetricity- the magnetic equivalent of electricity- was recently verified experimentally for the first time. Indeed, just as the stream of electric charges produces electric current, emergent magnetic monopoles have been observed to roam freely (generating magnetic current) in geometrically frustrated magnets known as spin ice. However, this is realized only by considering extreme physical conditions as a single crystal of spin ice has to be cooled to a temperature of $0.36 K$. Candidates to overcome this difficulty are artificial analogues of spin ice crystals, the so-called artificial spin ices. Here we show that, by tuning geometrical frustration down, a peculiar type of these artificial systems is an excellent candidate. We produce this material and experimentally observe the emergent monopoles; then, we calculate the effects of external magnetic fields, illustrating how to generate controlled magnetic currents. This potential nano-device for use in magnetronics can be practical even at room temperature and the relevant parameters (such as magnetic charge strength etc) for developing this technology can be tuned at will.
\end{abstract}

\maketitle

Geometric frustration occurs when the interactions among the constituents of a system that minimize energy can not be satisfied simultaneously in a local organizing principle due to geometric constraints. It plays important roles in several physical systems and particularly, it is behind the ground state degeneracy of spin ice compounds, which are so-called for their resemblance with the water ice behavior at zero temperature. Spin ice examples include $Dy_2Ti_2O_7$, $Ho_2Ti_2O_7$ and $Ho_2Sn_2O_7$ crystals which have been predicted\cite{Ryzhkin,CastelnovoNature,Jaubert} and experimentally observed\cite{3D-SI-EXP1,3D-SI-EXP2,3D-SI-EXP4} to support magnetic monopole-like excitations above the ground-state(s). Moreover, after managing to measure the amount of magnetic charge on these monopoles, some researchers were also able to inducing a magnetic flow analogues to electric current\cite{3D-SI-EXP3,Giblin11}. These motions and interactions of monopoles could develop technologies known as magnetricity and magnetronics. However, in spin ice crystals, these observations occur only at very low temperatures (around $0.36K$), which is a problem to construct magnetronic devices for direct applications in present days.

\begin{figure}
\centering
\includegraphics[width=8.5cm]{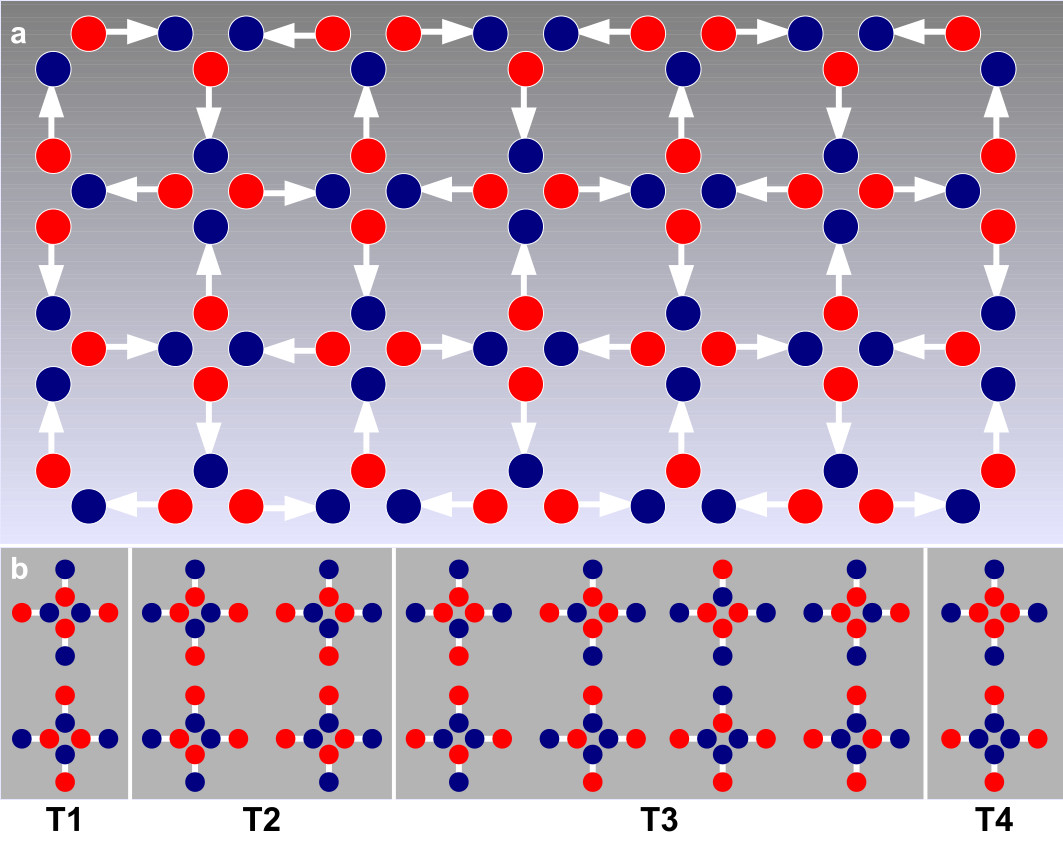}
\caption{ Illustration of the prototype material-by-design known as artificial spin ice. Dumbbell representation is used for characterizing the dipolar nature of the elongated magnetic nanoislands in an artificial square ice (red and blue circles indicate the two opposite poles of the dipoles). a) Configuration of the ground state with $4$ nanoislands meeting at each vertex. Here, all vertices obey the ice rule (with $two-in$, $two-out$ represented by $two-blue$, $two-red$). b) The $4$ possible vertex topologies $T_1$, $T_2$, $T_3$ and $T_4$ depicted in increasing energy from left to right; $T_1$ and $T_2$ do obey $2-in, 2-out$ ice rule. However, $T_2$ is more energetic than $T_{1}$ once it carries a net magnetization, while $T_3$ ($3-in, 1-out$ or $1-in, 3-out$) and $T_4$ ($4-in$ or $4-out$) violate ice rule and carry single and double net magnetic charge respectively. Note that all vertices in the ground state shown in \textbf{a} are in $T_{1}$.}
\end{figure}

\begin{figure*}[t]
\centering
\includegraphics[width=17.0cm]{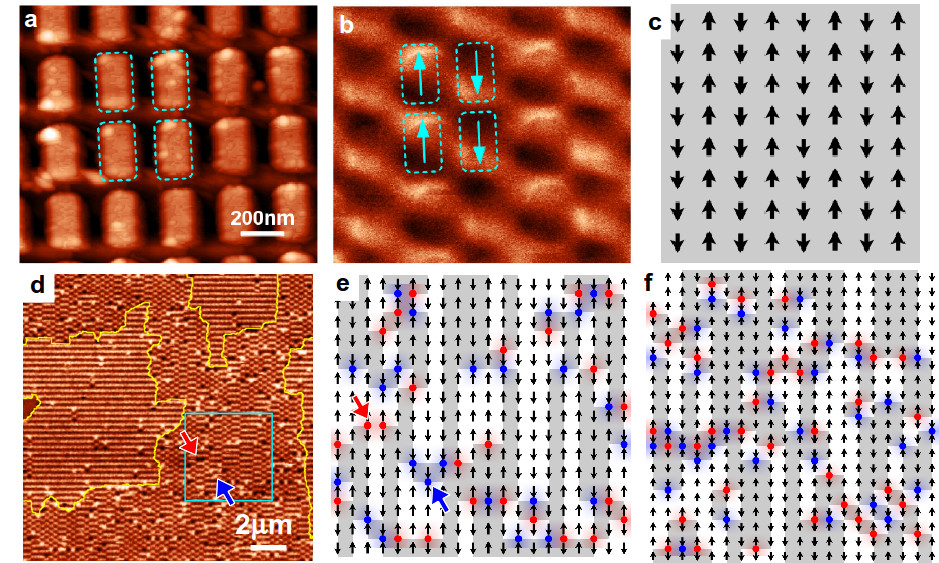}
\caption{ Unidirectional artificial spin ice. a) $AFM$ topography of a sample of the artificial spin ice device considered here. Cyan dots represent the position correspondent to each nanoisland. b), $MFM$ response of the device, depicting $4$ spins in the antiferromagnetic-like ground state. Arrows indicate the orientations of the dipoles.  c) Schematic view of the ground state configuration. d) $MFM$ measurement shows some ferromagnetic domains highlighted by yellow lines above the antiferromagnetic ground state. Several monopoles (black and white spots) spread for the large region of ground state. There are a concentration of monopoles in the edges of ferromagnetic domains. e) Amplified view of the region inside the cyan square highlighted in (d), with the nanoislands represented by black arrows. The blue and red circles are opposite monopoles above the ground state (grey regions) and along the boundaries of the ferromagnetic domains (white regions). Blue and red arrows shown here indicate the same monopoles inside the cyan square in (d), which are also indicated with similar arrows. f) Our theoretical results (obtained by point-like dipole and dumbbell models) always lead to domain configurations with pattern very similar to the experimental images as in (d) and (e).}
\end{figure*}

In principle, two-dimensional ($2d$) artificial analogues of spin ices (made by magnetic nanoislands\cite{Wang2006Nature} at room temperature) seem to be a natural alternative for spin ice crystals if one desires to construct magnetronic devices without requirement of extreme low temperatures and other inconveniences. Indeed, they have recently appeared in a number of lattice geometries like square \cite{Wang2006Nature}, hexagonal\cite{Branford-hexagonal,Hyderman-hexagonal}, triangular\cite{Mol-triangular1,Mol-triangular2} and rectangular\cite{FabioNJP2012}. In such sorts of $2d$ artificial spin ices ($ASI$'s), geometrical frustration also takes place and it is widely claimed to be one of the key elements behind their main physical properties, including the emergence of exotic types of magnetic monopoles connected by energetic strings\cite{RodrigoPRB}. For instance, in a square arrangement, where $4$ spins (magnetic moments) meet at each vertex, geometrical frustration yields ice rule which dictates  $2-in, 2-out$ spins at each vertex (Fig. $1a$). Energetic analysis shows that only $2$ of all possible $16$ arrangements at each vertex locally minimize the energy (topology $1$ or $T_{1}$ in Fig. $1b$). The remaining ones, even when satisfying ice rule (as topology $T_{2}$), cannot achieve energy minimization in this underlying geometry. Indeed, $T_{2}$ carries net magnetization while vertex topologies $T_{3}$ and $T_{4}$ break ice rule and carry net magnetic charge \cite{Mol2009JAP,Mol2010PRB,MarrowsNatPhys} (see Fig. $1b$). One of the main advantages of $2d$-$ASI$'s over its natural counterparts is that they may be fabricated with the desired parameters to yield precise physical properties, like monopole charge value, string paths and so forth. Even though monopoles and strings randomly appear as excitations above the ground-state of $ASI$’s, their locations and main properties may be relatively well determined\cite{Mol2009JAP,Mol2010PRB,MarrowsNatPhys}. However, the main challenge remains to know how to put such monopoles under a controlled motion providing magnetic current with desirable properties. A route for achieving this goal seems to be tuning string tension down to practically tensionless regime to allow free monopole displacement. At this extent, the usual $ASI$ arrangements considered so far do not collaborate\cite{Rodrigo12,Kapaklis12} and, therefore, the way would be to suitably design arrays in which their underlying geometries favor low string tension. Really, a manner of getting a substantial reduction of the string tension is to construct systems in which topologies obeying ice rule have the same energy. This has been shown to occur in some two-\cite{Hyderman-hexagonal,FabioNJP2012} and three-dimensional\cite{Mol2010PRB,Moessner2006,Moessner2009} $ASI$'s. However, release the string is not a sufficient condition to control the emergent monopoles motion and a magnetic current was never produced even in these degenerate artificial compounds. Here, both material and geometrical arrangement choices are in order to provide such a possibility.

\begin{figure}
\center
\includegraphics[width=8.5cm]{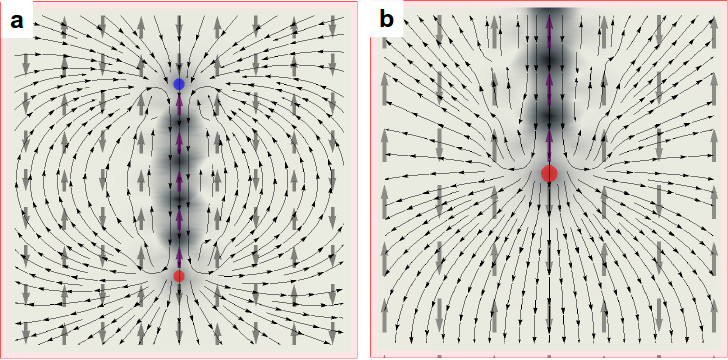}
\caption{Magnetic field lines for an excitation in the unidirectional artificial spin ice. a) The unidirectional artificial spin ice is represented by the large gray arrows in the background and the excitation consists of a pair of opposite monopoles (red and blue circles) separated by $5a$ and connected by a string. These opposite monopoles are always created along the $y$-direction and one lattice spacing apart, but they can be easily separated due to the low string tension. In addition, when an external magnetic field is applied from bottom to top, the red charge goes down while the blue one goes up and the poles become more and more distant. b)  Emphasis on the radial magnetic field lines around an individual monopole. }
\end{figure}

\begin{figure}
\centering
\includegraphics[width=8.5cm]{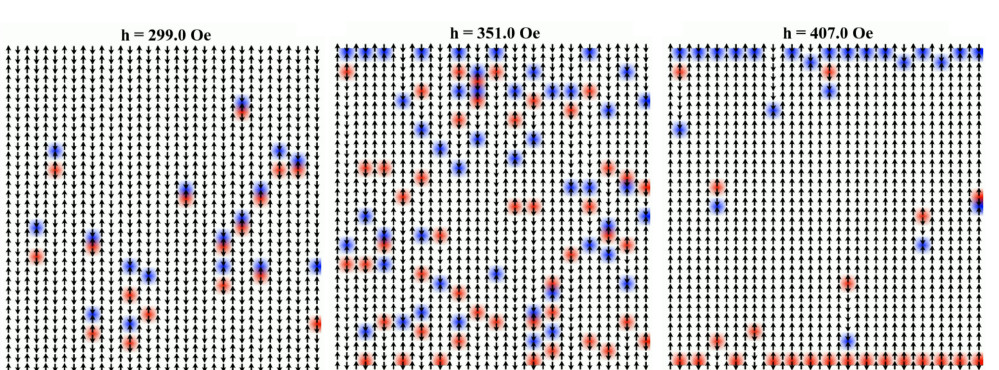}
\caption{Current of magnetic monopoles in the unidirectional artificial spin ice. An external magnetic field which increases with time, applied from bottom to top enlarges the ferromagnetic domains, providing the ordered motion of monopoles towards the sample borders, where they are eventually collected by suitable gates (not shown). With this field, the red charge type moves down while the blue type moves up. In the right panel, note the accumulation of blue charges in the top and red charges in the bottom. For more details about this magnetic current, see Supplementary Movies.}
\end{figure}

Accordingly, we report on the fabrication, characterization and theoretical analysis of a rectangular-type $ASI$ with nanoislands arranged to point along an unique direction (Fig.$2a$ shows $AFM$ image of our sample). Now, only $2$ dipoles meet at a given vertex and 'ice-like rule' is simply $1-in, 1-out$; the remaining $2-in$ or $2-out$ possibilities violate that rule and bear net magnetic charge. Therefore, there are only two topologies and once the ice rule is obeyed, energy minimization is locally achieved and no vertex geometrical frustration takes place. We shall show that excitations valued above the ground-state in this framework actually behave as localized magnetic monopoles interacting by means of a Coulombian plus a linear potential, the latter being physically represented by a (magnetized) string connecting the opposite charges, similarly to the scenario observed in usual $ASI$ systems. Nevertheless, here the string tension is too weak that monopole motion can be easily induced. The present framework is shown to be unique to easily localize magnetic monopoles (they show up at the borders of ferromagnetic domains), regulating them in an ordered and controllable motion, providing useful magnetic current at nanoscale.

In our samples, the elongated nanoislands (Fig. $2a$), with dimensions around $285 \times 165 \,{\rm nm^{2}}$ and bearing magnetic moment $\mu\approx 5.76\times 10^{-16} \, {\rm J/T}$, were obtained by electron beam nanolithography performed in $250 {\rm nm}$ polymethyl methacrylate ($PMMA$) electroresist, previously spin-coated on silicon. The islands are center-to-center spaced by around $a=280\, {\rm nm}$ and $b=350\,{\rm nm}$ (with ratio $a/b=0.8$) along $x$ and $y$ directions, respectively.  The final structures were obtained after deposition of $25\, {\rm nm}$ nickel thin film by electron beam evaporation in a pressure of $1.5\times 10^{-7}\, {\rm Torr}$ and lift-off process carried in acetone ultrasonic bath. Theoretical analysis is based upon two techniques: point-like dipole and dumbbell models. In the dumbbell model, each dipole has a finite length ($\delta$) and bears opposite magnetic charges ($\pm q_{m} $) at its end points (as $ \delta \to 0$, all results of the point-like dipole model is recovered). Main results do not depend on the model, although some quantitative deviations appear from each to other. In our calculations, the arrangements have linear sizes $(L_x, L_y)=(la, lb)$, so that the simulation takes into account $l^2$ dipoles (arrangements with distinct number of dipoles along $x$ and $y$ directions can be considered in the same way). To obtain the ground-state, we have used the simulated annealing process which is a Monte Carlo calculation where the temperature is slowly reduced to drive the system to a global minimum. Our Monte Carlo procedure consists of a single Metropolis algorithm. We have use open boundary conditions ($OBC$) and periodic boundary conditions ($PBC$) where $PBC$ were implemented by means of the Ewald summation. We have varied $l$ from $10$ to $70$ and our findings are practically independent of the choice of $PBC$ or $OBC$. The point-like dipole and dumbbell models were used. In the dumbbell model, assuming a rectangular geometry for the magnetic islands and using the nickel properties, the magnetic moment is $5.8 \times 10^{-16} A m^{2}$. The two orientations of the original dipole determine the distribution of charges and the length $\delta$ of the nanoisland gives the strength of the magnetic charges $q_{m}=\pm \mu/\delta$. Using the parameters of our experiments, the magnetic moment is $5.8 \times 10^{-16} A m^{2}$ and, therefore, the dumbbell model estimates the magnetic charge as $q_{m} \approx 2\times 10^{-9} A m$. Let us start by describing the system as their dipoles were point-like, so that its Hamiltonian reads

\begin{equation}\label{H-dipole}
H=D b^3 \sum_{i>j} \left[\frac{\vec{s}_i \cdot \vec{s}_j}{r^3_{ij}} -\frac{3 (\vec{s}_i \cdot \vec{r}_{ij})(\vec{s}_j \cdot \vec{r}_{ij})  }{r^5_{ij}} \right]\,
\end{equation}
where $D=\mu_0 \mu^2/ 4\pi b^3 $ is the dipolar constant, $b$ is the lattice spacing along $y$-direction (and the lattice spacing along the $x$-direction is $a \neq b$), $\mu$ is the magnetic moment of each island and $\vec{r}_{ij}$ is the vector between two distinct islands throughout the system. In addition, $\vec{s}_i\equiv \vec{\mu}_i/\mu$ is the normalized magnetic moment of each island pointing along positive or negative $y$-direction. The ground state theoretically found for this unidirectional rectangular artificial spin ice ($UDRASI$) is depicted in Fig.$2b$-$c$ , which is corroborated by $MFM$ experiments as shown in Fig.$2d$ (clearer and darker spots represent excited opposite magnetic poles). As may be easily realized, such a ground state is composed by lines of aligned dipoles with alternating magnetization, resembling an antiferromagnetic disposing. Figure $2e$ enlarges part of Fig. $2d$ with ferromagnetic domains and monopole pairs at their opposite borders along $y$-direction. Figure $2f$ shows how the theoretical calculations lead to spin patterns similar to that of experimental images.

The simplest excitation above the ground state consists in flipping one single dipole, so that one breaks $1-in, 1-out$ rule at two adjacent vertices. Additional flips along this line of dipoles (without further violation of the $1-in, 1-out$ rule) are performed to calculate the energy difference between the respective excited and ground states, as a function of the flipping steps size, $y=nb$ ($n$ is a positive integer). Such a difference is best-fitted by:
\begin{equation}\label{Energy-diff}
\Delta E= E_c + \beta y+ \frac{\cal{Q}}{y} ,
\end{equation}
where $E_c=7.2D$ is associate with the excitation energy creation, while $\beta=0.9 D/a$ is the string tension (Fig.$3a$). That such magnetic charges behave as point-like objects (Fig.$3b$) is further evidenced by the Coulombian interaction energy, ${\cal Q} / y$, with ${\cal Q}=-\mu_0 q^2_m/4\pi=-2.4Db$, from which the magnetic charge of each monopole reads $\pm q_m=\pm \sqrt{4\pi {\cal Q} / \mu_0}$, which typically goes around one hundred smaller than Dirac monopole charge, $q_{\rm Dirac}= 2\pi \hbar/\mu_0 e$. For the sake of comparison, usual square $ASI$ system presents $(E_c, \beta, {\cal Q})= (30D, 10D/b, -4Db)$ \cite{Mol2009JAP,Mol2010PRB,MarrowsNatPhys}. Thus, by tuning geometrical frustration down, $UDRASI$ drastically reduces the string tension, by around $10$ times, but keeping monopole charge in the same order of magnitude. In addition, while in usual $2d$ arrangements, strings have many possible paths to take from one monopole to its anti-monopole, yielding high entropic effect \cite{Mol2009JAP,Mol2010PRB,MarrowsNatPhys,Mol-triangular1,Mol-triangular2}, in $UDRASI$, it is restricted to be the shortest straight path connecting the pair along a given line of dipoles in the $y$-direction (Fig.$3a$).

In the dumbbell model, each charge $q_{m}= \pm \mu / \delta $ interacts with all the others charges of the system by a Coulombian potential and, therefore, the Hamiltonian becomes

\begin{equation}\label{Dumbbell}
H=\frac{\mu_{0}}{4\pi} \sum_{m>n} \frac{q_{m}q_{n}}{r_{mn}}.
\end{equation}
As expected, this alternative description provides the same ground-state and excitations of the point-like dipole method. The interaction monopole-antimonopole also obeys equation (\ref{Energy-diff}), but the constants are $(E_c, \beta, {\cal Q})= (16D, 0.05D/b, -2.5 Db)$. Although the creation energy together with monopole charge do not differ so much from that obtained with the point-like dipole model ($E_{c}= 7.2 D$, ${\cal Q} = -2.4 Db$), the string tension is about $18$ times smaller, corroborating the fact that in this system, monopoles are almost free to move.

Magnetic force microscopy ($MFM$) measurements strongly suggest $UDRASI$ systems as a very suitable framework to precise monopoles location. As shown in Fig.$2d$, they appear into pairs at the opposite borders of ferromagnetic domains, whose bulks comprise strings connecting them. Since magnetic monopoles have been precisely located throughout the system, we may eventually put them to move under a controlled way to achieve an useful magnetic current. This is perhaps the main advantage of $UDRASI$ over other typical $ASI$ systems: such a current is feasible by enlarging ferromagnetic domains, e.g., by application of an external magnetic field along the $y$-direction, yielding ordered and controlled motion of magnetic charges towards the borders of this artificial sample (see Fig.$4$ and Supplemental Material\cite{Movies}). By considering the excitation as shown in  Fig.$3a$, if the field is applied from bottom to top, the red charge type moves down while the blue charge type moves up; this occurs without further violation of $1-in, 1-out$ rule. There is a residual string tension but it is too weak to keep the monopoles together. In addition to induce monopoles motion along the $y$-direction, the field continually creates several pairs of them (Supplemental Material\cite{Movies}). Sometimes, blue and red types of charges meet due to their opposite paths, occurring the process of annihilation of monopoles. If the field increases with time, the monopole and magnetic current densities also increase. Since the charges move only unidirectionally, such an ordered motion comes to be a practical realization of {\em magnetricity} - a controlled current flow provided by magnetic rather than electric charge carriers in a nanoscaled circuit, leading to a sort of {\em magnetronics} device.\\

.
\begin{acknowledgments}
The authors thank CNPq, CAPES and FAPEMIG for
financial support.
\end{acknowledgments}

\end{document}